\documentclass[useAMS,usenatbib,a4paper]{mn2e}

\usepackage{graphics}
\usepackage{hyperref}
\usepackage{graphicx}
\usepackage{times}
\usepackage{amsmath}
\usepackage{natbib}
\usepackage{wasysym}
\usepackage[usenames]{color}
\usepackage{epsfig}
\usepackage{epstopdf}
\usepackage{subfigure}
\usepackage{graphics}
\usepackage{longtable}
\usepackage{amssymb}
\usepackage{latexsym} 
\usepackage{wrapfig}
\usepackage{mathrsfs}
\usepackage{morefloats}
\usepackage{bm}
\usepackage[T1]{fontenc}
\usepackage{aecompl}

\newcommand{\aap}{A\&A}
\newcommand{\apjl}{ApJ}
\newcommand{\apjs}{ApJS}

\newcommand{\ao}{Appl. Opt.}

\title[Palomar Kernel Phase Experiment]
{The Palomar Kernel Phase Experiment: Testing Kernel Phase Interferometry for Ground-based Astronomical Observations}
\author[B. J. S. Pope et al.]
{Benjamin Pope$^{1}$\thanks{E-mail: benjamin.pope@astro.ox.ac.uk}, Peter Tuthill$^{2}$, Sasha Hinkley$^{3}$, Michael J. Ireland$^{4}$,  \newauthor 
Alexandra Greenbaum$^{5}$, Alexey Latyshev$^{2}$, John D. Monnier$^{6}$, Frantz Martinache$^{7}$ \\ \\
$^{1}$Oxford Astrophysics, Denys Wilkinson Building, Keble Rd, University of Oxford, Oxford OX1 3RH, UK\\
$^{2}$Sydney Institute for Astronomy, School of Physics, University of 
Sydney, NSW 2006, Australia\\
$^{3}$School of Physics, University of Exeter, Stocker Road, Exeter EX4 4QL, UK\\
$^{4}$Research School of Astronomy \& Astrophysics, Australian National University, Canberra, ACT 2611, Australia\\
$^{5}$Department of Physics and Astronomy, The Johns Hopkins University, 3400 N. Charles Street, Baltimore, MD 21218\\
$^{6}$Department of Astronomy, University of Michigan, 1085 S University Ave, Ann Arbor, MI 48109-1090, USA\\
$^{7}$Laboratoire Lagrange, CNRS UMR 7293, Observatoire de la C\^{o}te d'Azur, Bd de l'Observatoire, 06304 Nice, France}

\begin{document}
\maketitle

\begin{abstract}
At present, the principal limitation on the resolution and contrast of astronomical imaging instruments comes from aberrations in the optical path, which may be imposed by the Earth's turbulent atmosphere or by variations in the alignment and shape of the telescope optics. These errors can be corrected physically, with active and adaptive optics, and in post-processing of the resulting image. A recently-developed adaptive optics post-processing technique, called kernel phase interferometry, uses linear combinations of phases that are self-calibrating with respect to small errors, with the goal of constructing observables that are robust against the residual optical aberrations in otherwise well-corrected imaging systems. Here we present a direct comparison between kernel phase and the more established competing techniques, aperture masking interferometry, point spread function (PSF) fitting and bispectral analysis. We resolve the $\alpha$ Ophiuchi binary system near periastron, using the Palomar 200-Inch Telescope. This is the first case in which kernel phase has been used with a full aperture to resolve a system close to the diffraction limit with ground-based extreme adaptive optics observations. Excellent agreement in astrometric quantities is found between kernel phase and masking, and kernel phase significantly outperforms PSF fitting and bispectral analysis, demonstrating its viability as an alternative to conventional non-redundant masking under appropriate conditions. 
\end{abstract}


\section{Introduction}
\label{intro}

Kernel phase interferometry \citep{2010ApJ...724..464M} is a powerful technique for image analysis, applicable to any observations which both Nyquist-sample all spatial frequencies, and have appropriately small wavefront errors. The method is based on a linear approximation, introduced first in \citet{2010ApJ...724..464M}, considering a transfer matrix propagating small phase errors in a pupil or redundant array into the corresponding space of $u, v$ baselines. The kernel of this matrix generates `kernel phases' that are a generalization of the self-calibrating closure phase quantity well-known in interferometry \citep{1958MNRAS.118..276J}. These linear combinations of phases have the property that small phase errors cancel, meaning for example that the residual aberrations after adaptive optics do not propagate at first order into the kernel phase measurements. The kernel phase technique holds promise for detecting objects at high contrast at or just inside the  diffraction limit $\sim \lambda/D$ of space telescopes, or of ground-based telescopes assisted with extreme adaptive optics (AO). Such instruments are now present on many of the world's largest optical telescopes, including PALM-3000 at Palomar \citep{2008SPIE.7015E..24B}, Subaru Coronagraphic Extreme AO (SCExAO) \citep{2009arXiv0905.0164M}, VLT-SPHERE \citep{2010lyot.confE..44B}, and the Gemini Planet Imager (GPI) \citep{2008SPIE.7015E..31M,2014PNAS..11112661M}. Moreover, the upcoming James Webb Space Telescope \citep{2006SSRv..123..485G} will be capable of nearly-diffraction-limited resolution at very high sensitivity, and it will be important to establish and benchmark the performance of kernel phase and other image analysis approaches to best exploit the high image quality it will provide. 

\citet{2013ApJ...767..110P} first applied kernel phase interferometry to a sample of brown dwarf systems imaged by \citet{2006AJ....132..891R} and \citet{2008AJ....135..580R} using the \emph{HST}-NICMOS NIC1 camera. Where the original papers found a total of ten binary systems out of 72 studied, \citet{2013ApJ...767..110P} recovered all of these and found five additional systems with high confidence, as well as four other new more marginal candidates at lower confidence or higher contrast. 

Kernel phase interferometry has not previously been used to resolve a close system with ground-based full-aperture extreme adaptive optics observations, and a detailed discussion of kernel phase reduction compared to other competing techniques has not been published. \citet{2011SPIE.8151E..33M} reported the extraction of kernel phases from a Keck~II NIRC image, but did not report the detection of any companion. \citet{2014IAUS..299..199I} have presented a kernel-phase image reconstruction of the LkCa15 system in an M filter. In this paper, we discuss kernel phase and NRM observations of the close binary system $\alpha$~Ophiuchi ($\alpha$ Oph, or Rasalhague) under identical observing conditions on the same night, with a view to using this as a benchmark for comparing the two methods. This is the first time such a simultaneous comparison has been made. A preliminary analysis of these observations was presented in conference proceedings by \citet{2013aoel.confE...6M,2015AAS...22531302H}.

\citet{2011ApJ...726..104H} carried out an adaptive optics non-redundant masking (NRM) study of this system, a nearby binary with an A5 III primary, in order to characterise its orbital parameters. \citet{2011ApJ...726..104H} find a contrast ratio of $27.9 \pm 8.3$ in the K band from resolved Palomar-PHARO \citep{2001PASP..113..105H} imaging. The primary is known to be rotating at $\sim 89\%$ of its predicted breakup velocity \citep{2009ApJ...701..209Z}, and asterosesimic analysis with MOST shows rotationally-modulated $g$-modes that probe the conditions of the interior \citep{2010ApJ...725.1192M}. Establishing its mass with precision is therefore valuable for constraining models of its rotational dynamics. \citet{2011ApJ...726..104H} predicted the companion would pass periastron at $\sim 19$ April 2012, and for this reason there was an observing campaign in 2012 to track its orbit at its apparent closest point, which is particularly critical in delivering a fully-constrained dynamical orbit. While $\alpha$ Oph is ordinarily a well-separated binary, at periastron the companion is buried within the PSF of the primary. 

\subsection{Observations}
\label{observations}

Observations were made on 26 June 2012, two months after periastron, using the PHARO camera on the 5.1~m Hale Telescope at Mt.~Palomar Observatory. Data were obtained using a 9-hole aperture mask, an 18-hole aperture mask and with the full aperture (with no mask) in CH$_4$ and $K_s$ bands, whose filters are centred at 1.57 $\mu$m and 2.145 $\mu$m respectively with bandpasses of 0.1 and 0.310 $\mu$m. The 9-hole mask contained projected baselines ranging from 0.75 to 4.15~m with a projected hole diameter of 0.5~m, while for the 18-hole mask the projected baselines ranged from 0.37 to 4.81~m with a projected hole diameter of 0.25~m.

In addition to $\alpha$ Oph, PSF reference stars $\epsilon$ Oph and $\epsilon$ Her were observed as calibrators. These data were then reduced to a standard FITS cube form using existing masking code. Regrettably, the full aperture observations in the CH$_4$ band suffered from poor AO performance and detector saturation and were excluded from the present study. The seeing varied between 1.5 and 2 arcseconds during the observations. By modelling the PSF, we determined the median Strehl across the exposures to have been $\sim 0.51$ in the $K_s$ band. 

In the full-pupil imaging, a neutral-density filter was used to diminish the brightness of the star, as is necessary to avoid saturating the science camera. This introduces a `ghost' (a reflection artefact), which severely limits the maximum size of the window that can be used in kernel phase analysis, as discussed in Section~\ref{comparison}. In future, it would be beneficial to choose filters in such a way as to avoid ghosts wherever possible, or to avoid using such filters altogether, for example by using very short exposure times or simply observing fainter targets.

\begin{figure}
\center
\includegraphics[width=0.7\textwidth]{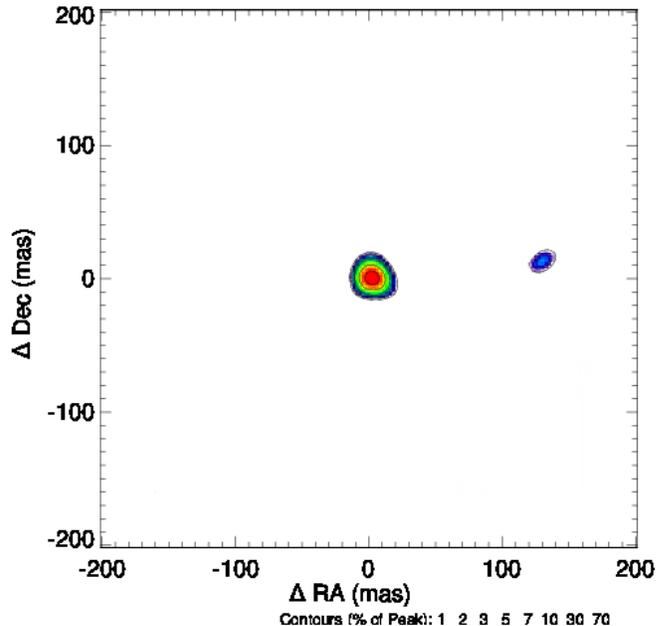}
\caption{Maximum Entropy model-independent image reconstruction using \textsc{BSMEM} for $K_s$ band non-redundant masking data.}
\label{Ksmap}
\end{figure}

Non-redundant 18-hole aperture masking observations with a Br$\gamma$ filter were processed to obtain the arguments of the mean bispectrum, i.e. bispectral-amplitude-weighted average closure phases. We used a Markov Chain Monte Carlo (MCMC) method to fit a binary model to these closure phases, recovering a companion at $131.2\pm1.4$ mas separation, position angle $82.8 \pm0.7^{\circ}$ and $27.6 \pm 1.2$ contrast. As we were not able to obtain Br$\gamma$ observations with a filled aperture, these are useful only for comparison with the kernel phase data. 
Non-redundant 9-hole observations with a $K_s$ filter found the same companion at $129.6 \pm 2.2$ mas separation, $83.5\pm1.1^{\circ}$ position angle and $28.7 \pm 2.3$ contrast. This contrast ratio is consistent with the $27.9 \pm 8.3$ in the K band reported in \citep{2001PASP..113..105H}. In addition to this, we performed a maximum entropy model-independent image reconstruction using \textsc{BSMEM} (Figure~\ref{Ksmap}), in which it is apparent that the parametric model accurately captures the information about the source intensity distribution. These aperture masking observations set the standard with which the kernel phase-based analyses must be compared.

\section{Processing and Results}
\label{palomarresults}

In Sections~\ref{calibs} and~\ref{bayes} we discuss the methods used to analyse data in this paper. In the interests of reproducibility and open science, we have made public our code and data: \textsc{.fits} files for our raw data are stored on Figshare \footnote{\url{http://figshare.com/s/4e69f7b2b30411e4bf4a06ec4bbcf141}}, together with the IPython notebook which was used to analyse these observations; and the \textsc{pysco} PYthon Self-Calibrating Observables package, a Python module for extracting and analyse kernel phase data, is available in a public GitHub repository \footnote{\url{https://github.com/benjaminpope/pysco}}. All other packages used in this analysis are publicly distributed elsewhere. We welcome efforts by other researchers to apply this body of analysis software to these or other data, with appropriate citation.

In all full-frame images, the peak of the PSF pushed the detector into its nonlinear response regime. Kernel phase requires strict linearity, being a Fourier technique, and so each frame therefore had to be calibrated with a nonlinear gain curve map in order to restore linearity.  The core of $\epsilon$ Oph was fully saturated in $K_s$ band, and therefore $\epsilon$ Her was used as the sole calibrator. 

\subsection{Kernel Phase Extraction and Calibration}
\label{calibs}

As discussed in Section~\ref{intro}, kernel phases are self-calibrating linear combinations of phases, which are robust with respect to small residual wavefront errors. We calculate a matrix to generate these \emph{a~priori} based on an assumed discrete model of the pupil.

We obtained direct images of the PHARO `medium cross' pupil, and found the ratio of the outer radius to the central obscuration, and of this radius to the thickness of the spiders, to differ to a small extent from the nominal values reported in \citet{2001PASP..113..105H}. This is very important to establish carefully, as information from longer baselines than exist in the telescope consists purely of noise and will corrupt any kernel phases obtained. In order to establish a precise pupil model as is necessary for kernel phase, we used the visibility amplitudes extracted from the point-source calibrator, $\epsilon$ Her, to constrain the overall scale of the pupil. This may differ from the published values because of an error in the measured projected pupil size, or an offset in the effective filter bandpass, which we model as the nominal $K_s$ band centre of $2.145~\mu$m, but may differ from this nominal value due for example to a slope in the stellar spectrum. 

\begin{figure}[h]
\center
\includegraphics[width=0.55\textwidth]{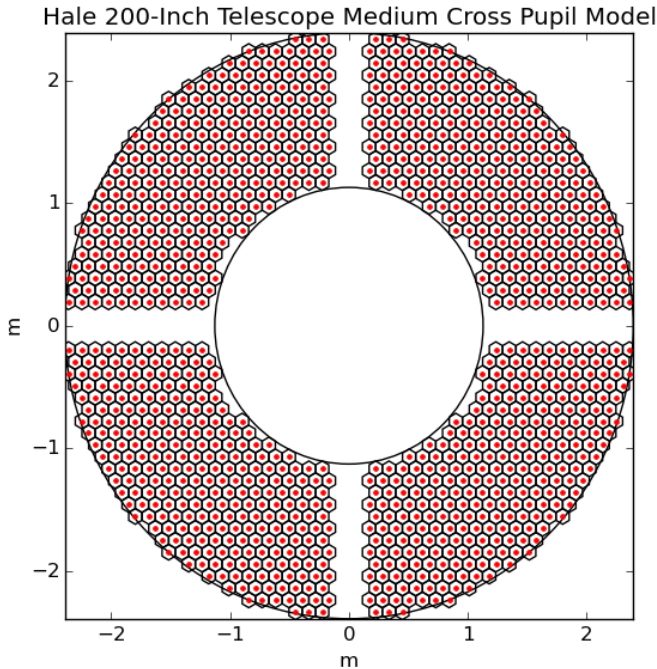}
\caption{Hale Telescope medium-cross pupil model. Red dots represent pupil sampling points; note that they avoid the spiders, which on the Hale Telescope are vertical and horizontal with respect to the detector axes.}
\label{hale}
\end{figure}

The absolute magnitude of the Fourier transform of the image of a point source, in the case of a flat wavefront, is a map of the optical modulation transfer function. This is itself found as the autocorrelation of the pupil, whose magnitude is approximately given by the redundancy of each baseline in a discrete pupil model. We therefore varied this overall outer scale in the vicinity of the value reported in \citet{2001PASP..113..105H}, which lists a projected radius of 2.32~m. In order to be as sensitive as possible to the outer radius, we perform a least-squares fit by brute force between the logarithms of the redundancy matrix elements and the magnitude of the stacked Fourier transforms of all 100~observed frames. The fit is best-conditioned by the low visibilities, at the edge of the pupil, and there is some discrepancy at intermediate visibilities (low spatial frequencies), where we are sensitive to the faint binary-like signal of the ND filter ghost and residual low-order aberrations. Ideally, we would model this pupil cojointly with the binary model, and marginalize over uncertainties, which in the present circumstances we are unable to do due to the prohibitively long computational times. The best fit is found with an outer projected radius of 2.392~m, which we therefore adopt as fixed in the following analysis. The discretized pupil generated with this model is shown in Figure~\ref{hale}, containing 1128 elements, and generating 3256 baselines. 

Using a singular value decomposition, we find this model to generate 2692 kernel phases. We centre each image in real space and recentre it to sub-pixel precision by subtracting a phase slope in its Fourier transform, and apply this matrix to phases extracted from the corresponding 3256 baselines in this Fourier transform in order to obtain the kernel phases.

There are several differences between the application of kernel phase to this dataset and to the previously-published \emph{HST} sample in \citet{2013ApJ...767..110P}. In particular, each observation consists of a datacube of 100 frames, yielding excellent experimental diversity so that the statistics on each kernel phase can be readily recovered, as opposed to the case with the \emph{HST} snapshot data where an ensemble average over many different targets was required. Kernel phases are therefore extracted separately from each frame of data, and then combined such that in the following Sections we take as our data the ensemble mean of each kernel phase over the set of frames, and the statistical uncertainties are taken to be the standard errors of the mean (SEM). 

As is standard practice in NRM interferometry, the PSF reference stars $\epsilon$ Oph and $\epsilon$ Her were processed in the same way. By subtracting the kernel phases measured on these point sources, it is therefore possible to calibrate systematic offsets in the instrumental kernel phase measurements. The uncertainties on each of the calibrator's kernel phases, again taken to be the standard error of the mean, are added in quadrature to the uncertainties on the science target's kernel phases.

In addition to this, we also add in quadrature a second error term of $1.35^\circ$ to account for uncalibrated systematic errors. We fit a parametric binary model to the data as described in Section~\ref{bayes}, and iteratively adjust the magnitude of this additional error term so that the reduced $\chi^2$ of the best-fitting parameters is approximately unity.

\subsection{Bayesian Parameter Estimation}
\label{bayes}

The next steep is to fit a parametric model to these kernel phase data, defined by the binary parameters separation (mas), position angle (deg) and contrast, proceeding in a similar fashion to \citet{2013ApJ...767..110P}. We estimate these parameters using two Bayesian inference algorithms, namely \textsc{MultiNest} \citep{2009MNRAS.398.1601F}, an implementation of multi-modal nested sampling, and \textsc{emcee} \citep{2013PASP..125..306F}, an affine-invariant ensemble Markov Chain Monte Carlo sampler. We used both approaches firstly as a check for consistency, but also because they have complementary strengths \citep{2014MNRAS.437.3918A}: on the one hand, \textsc{MultiNest} efficiently and reliably converges on the global peak of a posterior distribution without significant sensitivity to an initial guess and avoids being trapped in local likelihood maxima. On the other hand, \textsc{emcee} is more effective at exploring and characterizing potentially non-Gaussian, curving degeneracies in the shape of the posterior mode; as noted in \citet{2013ApJ...767..110P}, there is typically significant degeneracy between separation and contrast in kernel phase fits to systems at close to the diffraction limit, and it is important to explore the shape of this curve.

\begin{figure}
\center
\includegraphics[scale=0.4]{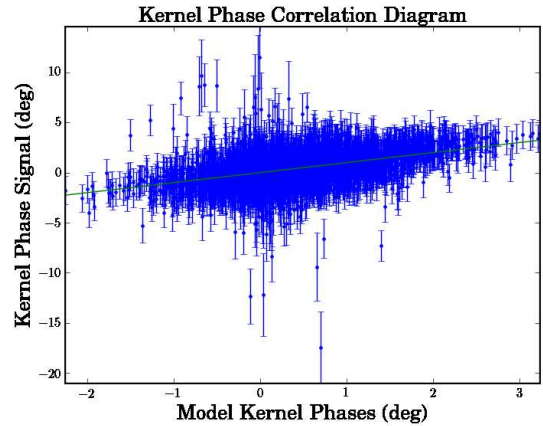}
\caption{Correlation diagram for $\alpha$ Ophiuchi kernel phases in $K_s$ band. We plot model kernel phases on the $x$-axis and the observed signal on the $y$-axis, such that for a good fit we expect the data to lie on a straight line of gradient unity (overplotted green line).}
\label{alpophcorr}
\end{figure}

We began by running \textsc{MultiNest}, obtaining the parameter estimates listed in Table~\ref{palomartable}. The corresponding correlation diagram is displayed in Figure~\ref{alpophcorr}. After our first attempt with no additional uncertainty added in quadrature, we iteratively re-ran the MCMC adding an additional error term in quadrature until the fit of the posterior mean achieved a reduced $\chi^2$ of approximately unity. This term was found to be $\sim 1.35^\circ$ in the kernel phase case, and $7.0^\circ$ in the case of the bispectral phases.  

As discussed above, interferometric determinations of binary parameters at close to the diffraction limit often suffer from degeneracy between contrast and separation. As a result, we used the \textsc{MultiNest} output to initialize an \textsc{emcee} run with 100 walkers and 200 burn-in steps and recorded 1000 subsequent steps to sample from the posterior. From this, it is apparent that there is only a small degree of covariance between these parameters, and we find good agreement between the \textsc{MultiNest} and \textsc{emcee} estimates of the posterior mean and standard deviation. 

In order to test whether the kernel phase processing itself introduces a bias into the contrast and separation estimates, we simulated binaries with the same parameters each as the best fit to the kernel phase full aperture observations and to the non-redundant 9 hole closure phase measurements. These simulations use no atmosphere, but include a realistic Palomar `medium cross' pupil model identical to that used to derive the kernel phase relations. Model fitting is performed with \textsc{MultiNest} as in Section~\ref{bayes}, with uncertainties on each kernel phase taken to be the same as from the real observations. For an input model with the parameters of the kernel phase model (129 mas, 83.6 degrees, 34.2 contrast) we retrieve $129.4 \pm 1.2$ mas separation, $83.5 \pm 0.3$ deg position angle and contrast $34.3 \pm 1.0$; and for the aperture masking parameters (129.6 mas, 83.5, 28.7 contrast), we retrieve $129.7 \pm 1.0$ mas, $83.5 \pm 0.3$ deg and $28.8 \pm 0.8$ contrast. It is evident from these fits that kernel phase fitting itself introduces no bias towards lower contrasts. 

\begin{table}
  \caption{Binary Parameter Estimates for $\alpha$ Ophiuchi at JD 2456104.847025.}
  \label{palomartable}
  \begin{center}
    \leavevmode
    \begin{tabular}{llll} \hline \hline              
  Mode & Separation          & Position Angle              & Contrast      \\ 
  &(mas) & (deg)             & ($K_s$)  \\ \hline 
  Kernel Phase & 129.3 $\pm$ 1.2  & 83.6 $\pm$ 0.3  & 34.2 $\pm$ 1.1 \\
  PSF Fitting & 131.9 $\pm$ 0.5 & 84.3 $\pm$ 0.4 & 19 $\pm$ 2.4 \\ 
  Aperture Masking & 129.6 $\pm$ 2.2 & 83.5 $\pm$ 1.1 & 28.7 $\pm$ 2.3 \\
  Bispectral Phase & 140.9 $\pm$ 1.0 & 86.7 $\pm$ 0.7 & 15.4 $\pm$ 0.7 \\

    \end{tabular}
  \end{center}
\end{table}

\subsection{Comparison with Filled-Aperture Bispectral (Closure) Phase Analysis}
\label{bispectrum}

Using the same full-aperture observations analysed with kernel phase in Section~\ref{bayes}, we also performed an identical analysis using the arguments of the bispectrum, i.e. closure phases. The closure phases are the arguments of the bispectrum or `triple product' of three complex visibilities around a closing triangle, and we therefore call these closure phases or bispectral phases interchangeably. \citet{1986OptCo..60..145R} recognized bispectral analysis as being equivalent to the existing triple-correlation method of speckle masking \citep{1970A&A.....6...85L,1977OptCo..21...55W}, and it provides a more robust observable than raw phases even on partially redundant pupils \citep{1992JOSAA...9..203H}. 

For a redundant pupil, there are a combinatorically large number of baselines, and for reasons of hardware memory we are not able to use a pupil model as dense as in the above kernel phase analysis. We instead use a pupil model with the same dimensions but only 508 pupil samples. Using this coarser model we conduct a kernel phase fit, finding good agreement with the denser model, with a separation of $127 \pm 1.8$ mas, a position angle of $83.2 \pm 0.4$ degrees, and a contrast ratio of $32.2 \pm 1.4$. We see that this is in reasonable agreement with the denser model and aperture masking observations. 

We then find all possible combinations of triangles and test for closure, finding 378662 closing triangles in our redundant pupil model. As there are only 1456 independent $u, v$ baselines, the information in the raw bispectrum is extremely redundant, and unless we model our data in a reduced-dimensionality representation, we will both encounter unnecessary computational cost, and underestimate our uncertainties. We therefore first construct the $378662 \times 1456$ matrix containing the full set of closure relations, and find a rank-reduced operator with the same range using a sparse SVD. Using this, we find that the space of closure phases is spanned by the expected $N_{baseline} -2 = 1454$ orthonormal vectors, and use these as our bispectral observables. These are therefore linearly-independent closure phases \citep{2015ApJ...801...85S}, but we do not have a sufficiently large number of observations to re-diagonalize these as statistically-independent closure phases as in \citet{2012ApJ...745....5K,2013MNRAS.433.1718I}. It is important to note that in the general non-redundant cases these orthonormal closure phases do not span the same space as kernel phases, and that only in the case of a non-redundant pupil are these two spaces of observables expected to be the same. Ideally, we would for each triangle average the complex bispectrum across all frames, extract the phase of the resulting mean complex bispectrum, and then project these onto the minimal spanning set of orthonormal closure phases. Due to the combinatorically large number of triangles this is not possible, and we instead average the orthonormal closure phases themselves. 

Data are processed as for the kernel phases in Section~\ref{calibs}, except using this orthonormalized matrix of closure phase relations instead of the kernel phase matrix. We recover the binary at a separation of 140.9 $\pm$ 1.0 mas, 86.7 $\pm$ 0.7 degree position angle and a contrast of 15.4 $\pm$ 0.7. The very small uncertainty quoted on this fit is statistical, and clearly the dominant error here is systematic. The position angle and separation are roughly similar to that determined in Section~\ref{bayes} with kernel phase, but offset by $\sim 10$~mas in separation, and the best-fitting contrast is much lower, at 15.4 $\pm$ 0.7, which is $\sim 2$ times lower than the best kernel phase or masking estimate and similar to that from PSF fitting. This is consistent with the effects of a speckle introduced by a phase aberration perturbing all the full-frame images, while kernel phase and aperture masking are by construction resilient against this form of aberration.

While the difference in sampling density means that we do not compare the kernel phase and bispectral methods on a level playing field, we note that finding the full set of triangles generated by the denser model was not possible due to memory constraints, and therefore the bispectral method is inherently more limited than kernel phase in its applicability to very dense, redundant pupils. We are therefore restricting our comparison to methods of equivalent computational resource usage.

\begin{figure}
\center
\includegraphics[scale=0.4]{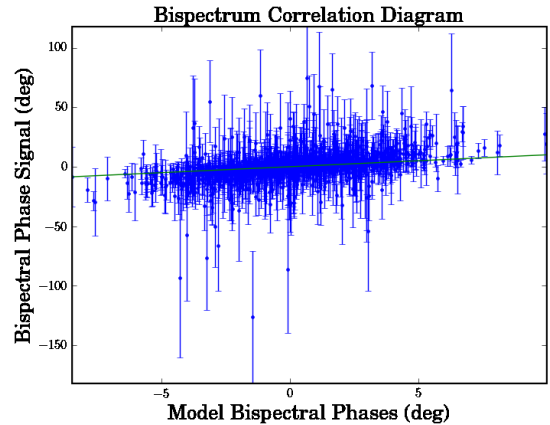}
\caption{Correlation diagram for bispectral phases for $\alpha$ Ophiuchi in $K_s$ band. We plot model phases on the $x$-axis and the observed signal on the $y$-axis, such that for a good fit we expect the data to lie on a straight line of gradient unity.}
\label{bispeccorr}
\end{figure}

\subsection{Comparison with PSF Fitting}
\label{psffitting}

In this situation we lack the diversity of calibration images to use the most advanced PSF analysis techniques, such as LOCI \citep{2007ApJ...660..770L} or KLIP \citep{2012ApJ...755L..28S}, but the 100 frames of our single calibrator are sufficiently diverse to permit PSF fitting. We adopt a maximum-likelihood approach, finding the best-fitting incoherent sum of two shifted calibrator PSFs that minimize a $\chi^2$ objective function for each individual frame of $\alpha$ Oph. The offset and scaling between these two calibrator PSFs is then taken to be the model parameters of the binary system. In Table~\ref{palomartable} we report the mean and standard error of the mean of all such fits with a $\chi^2$ within a factor of 2 of the median $\chi^2$ across all frames, to allow for the possibility of failing to retrieve the binary in some frames.

We note the low contrast found in the results of PSF fitting, not found in the kernel phase analysis. We suggest that a quasi-static speckle very close to the position of the binary companion, or a variable incoherent second ghost at the same location, could cause this effect, which is corrected in the kernel phase approach. This good kernel phase correction is consistent with this being a phase-aberration induced speckle; in the case of an amplitude-induced speckle, it would be desirable to find an equivalent `kernel amplitude' which would be self-calibrating with respect to such an aberration, but such a quantity is not presently known. 

We note that this PSF fitting approach is enabled by the diversity over both many calibrators and many frames on each calibrator, whereas in principle any single image can be analysed with kernel phase given appropriate single images of calibrators. This is an advantage for kernel phase in campaigns where many short exposures are not feasible, or in space telescope snapshot campaigns, where PSF diversity may otherwise be lacking.

\section{Comparison between Masking and Kernel Phase}
\label{comparison}

For our benchmark $\alpha$ Oph dataset, the kernel phase analysis strongly favours a binary model over a point source, and successfully obtains the same system configuration as found with aperture masking. The results are displayed in Table~\ref{palomartable}. The spatial astrometric components agree remarkably well, to much better than $1\sigma$, and in contrast the kernel phase is slightly higher, at 34.2 vs 28.7. As the uncertainties on each are $\sim 1.1$ and $\sim 2.3$ respectively, we see these estimates do not entirely agree, though the higher contrast value derived from kernel phase may represent a systematic error which may require further improvements to the method. As noted below, the uncertainties on the kernel phase-derived contrast are probably underestimated here, though with present software implementations and the limited number of frames available in these observations, this cannot be resolved at present. Nevertheless, it is clear that kernel phase presents a realistic alternative to aperture masking for telescopes with extreme AO, with the potential for significant advantages in throughput and Fourier coverage. 

The science target was observed in the middle of the detector, near the corner of the four CMOS chips which tile the focal plane. This leads to lines of noise running through the sides of the PSF. On the other hand, the calibrators were observed in the middle of each chip (`dithered'), as is standard practice. The subtraction of systematic errors from the science data was therefore imperfect, and in future it will be important to dither the science source and calibrators identically.

The binary astrometry from kernel phase is reasonably close ($\sim 1.5-2\sigma_{masking}$) to that obtained by aperture masking; this is remarkably good agreement given that the observations were performed under far from ideal conditions: as noted in Section~\ref{observations}, on the night these observations were made, the seeing was $1.5-2$ arcseconds, which is substantially worse than median for Palomar. This translated to the relatively low Strehl of 0.5 in $K_s$, somewhat low for extreme-AO and toward the lower end of the kernel phase regime. We therefore note that we may expect improved performance under better AO conditions in future.

The retrieved binary parameters are remarkably insensitive to the size of the super-Gaussian window used in preprocessing. By the convolution theorem, windowing an image is equivalent to convolving its Fourier transform with a kernel whose dimension is inversely proportional to that of the window. A narrower window therefore has a wider convolution kernel, which blurs Fourier phase information. This results in both a blurring in the Fourier plane, which is especially significant at high spatial frequencies where this mixes real signals from inside the support of the modulation transfer function with noise from outside its support.

In addition to the above instrumental errors, we noted in Section~\ref{psffitting} that systematic optical aberrations are likely to remain even after calibration. By using an ensemble of calibrators as described in \citet{2012ApJ...745....5K}, it is possible to substantially improve the correction of systematic errors. Residual uncalibrated systematics enter at first order in phase, and third order in kernel phase, and can introduce statistical correlations between kernel phase relations which are algebraically orthogonal. It is possible to diagonalize kernel phases and closure phases in a Karhunen-Lo\`{e}ve basis which properly takes into account their statistical covariance \citep{2013arXiv1301.6205I}; such a calibration is improved significantly with greater calibrator diversity than is available in this work. In future, it will be valuable to include more calibrator sources, and observe these and the science target with a range of pupil orientations to maximize the calibration diversity. We expect that this may reduce systematic effects, and also increase the parameter uncertainties somewhat, given that presently we assume data to be independent which are in fact correlated.

The issue of pupil modelling remains an outstanding problem for kernel phase, in that at present it is not feasible to marginalize over uncertainties in pupil scale, or to model the pupil as densely as may be preferred for high-performance applications. There are therefore systematic errors associated with any mismatch between the discrete pupil model and the real effective model of the telescope, as well as any amplitude aberrations or spatial variations in transmission. Resolving this difficulty is beyond the scope of this paper, but will be important for future work.

\section{Discussion}
\label{refdisc}

The recovery of the $\alpha$~Oph binary system illustrates both the potential of kernel phase in conjunction with extreme AO, and the potential for improvements in future observations. It is clear that kernel phase recovers the binary parameters with remarkable precision, and it will be a valuable tool in probing systems than cannot be observed with aperture masking. In this test case, under ideal conditions for aperture masking and more challenging conditions for kernel phase, nevertheless kernel phase delivers comparable results. The performance of kernel phase in this low to moderate Strehl, single calibrator regime is not expected to be representative of higher Strehls and multiple calibrators - an analysis of which would be important future work.

We have also demonstrated the benefits of using kernel phases over more standard PSF fitting and bispectrum (closure phase) approaches in parameter estimation. Deconvolving structure from an AO-corrected PSF is significantly enhanced by the use of kernel phase, and we expect this will enable new science to be done at and near the diffraction limit. 

As noted in \citet{2013ApJ...767..110P}, wavelength diversity across several filters can help alleviate the degeneracy between separation and contrast, by jointly fitting to kernel phases extracted in several bands and enforcing the condition that the position of a companion must be fixed, while its flux can vary. This is a promising option for future kernel phase work, as the extreme AO systems SPHERE and GPI, as well as the P1640 instrument on Palomar, are equipped with integral field spectrographs which can obtain images in many wavelength channels simultaneously. 

Given these encouraging results, we see that the best current adaptive optics systems are already able to make use of kernel phase for high contrast imaging. In particular, we have shown that in the extreme-AO regime, kernel phase obtains comparable results to those using non-redundant masking. Where for hardware reasons or due to throughput considerations it is not possible to use a mask, or where very dense Fourier coverage is desired for imaging, the kernel phase approach may be much more effective than standard alternatives, opening up new parameter space for high-resolution imaging of faint companions and circumstellar environments. We have also discussed observing strategies, and in particular the importance of calibrator sources and wavelength diversity, which will be of use in planning future kernel phase work from the ground.

\section*{Acknowledgements}
\label{acknowledgements}
This research has made use of NASA's Astrophysics Data System. This research has also made use of Astropy, a community-developed core \textsc{Python} package for Astronomy \citep{2013A&A...558A..33A}, the \textsc{IPython} package \citep{PER_GRA:2007}, and \textsc{matplotlib}, a \textsc{Python} library for publication quality graphics \citep{Hunter:2007}. 

BP thanks Balliol College and the Clarendon Fund for their financial support for this work.

This work was performed in part under contract with the California Institute of Technology, funded by NASA through the Sagan Fellowship Program.

We are grateful to Anthony Cheetham, Rupert Allison, Michael Bottom, Matthew Kenworthy, Daniel Foreman-Mackey, Patrick Roche and Suzanne Aigrain for their helpful comments. We would especially like to thank our referee, Chris Haniff, for his advice in improving this paper.

\bibliographystyle{aa}
\bibliography{ms}

\end{document}